\documentclass[conference]{IEEEtran}
\IEEEoverridecommandlockouts
\usepackage{subfigure}
\usepackage{mathtools}
\usepackage{cite}
\usepackage{amsmath,amssymb,amsfonts}
\usepackage{algorithmic}
\usepackage{graphicx}
\usepackage{textcomp}
\usepackage{xcolor}
\usepackage{orcidlink}
\usepackage{tikz}
\usepackage{makecell}
\usepackage{floatrow}
\usetikzlibrary{positioning}

\usepackage{etoolbox}
\makeatletter
\patchcmd{\@makecaption}
  {\scshape}
  {}
  {}
  {}
\makeatletter
\patchcmd{\@makecaption}
  {\\}
  {.\ }
  {}
  {}
\makeatother
\def\tablename{Table}

\newif\ifdraft
\drafttrue
\ifdraft

\else
\newcommand{\note}[1]{}

\fi

\def\BibTeX{{\rm B\kern-.05em{\sc i\kern-.025em b}\kern-.08em
    T\kern-.1667em\lower.7ex\hbox{E}\kern-.125emX}}
\begin{document}

\title{\huge{Solving a real-world modular logistic scheduling problem with a quantum-classical metaheuristics}}

\author{
\IEEEauthorblockN{Florian Krellner$^*$}
\IEEEauthorblockA{\textit{SAP SE} \\
Walldorf, Germany\\
florian.krellner@sap.com} \\
\IEEEauthorblockN{Sarah Braun}
\IEEEauthorblockA{\textit{Siemens AG} \\
Munich, Germany\\
sarah.braun@siemens.com}
\and
\IEEEauthorblockN{Abhishek Awasthi$^*$}
\IEEEauthorblockA{\textit{BASF Digital Solutions GmbH}\\
Ludwigshafen am Rhein, Germany\\
abhishek.awasthi@basf.com}\\
\IEEEauthorblockN{Michael Poppel}
\IEEEauthorblockA{\textit{Aqarios GmbH} \\
Munich, Germany \\
michael.poppel@aqarios.com}
\and
\IEEEauthorblockN{Nico Kraus}
\IEEEauthorblockA{\textit{Aqarios GmbH} \\
Munich, Germany\\
nico.kraus@aqarios.com} \\
\IEEEauthorblockN{Daniel Porawski}
\IEEEauthorblockA{\textit{Aqarios GmbH}\\
Munich, Germany\\
daniel.porawski@aqarios.com}
    \thanks{$^*$ Corresponding authors}
    \thanks{ }
    \thanks{© 2025 IEEE. Personal use of this material is permitted. Permission from IEEE must be obtained for all other uses, in any current or future media, including reprinting/republishing this material for advertising or promotional purposes, creating new collective works, for resale or redistribution to servers or lists, or reuse of any copyrighted component of this work in other works.}
}

\maketitle

\begin{abstract}
This study evaluates the performance of a quantum-classical metaheuristic and a traditional classical mathematical programming solver, applied to two mathematical optimization models for an industry-relevant scheduling problem with autonomous guided vehicles (AGVs). The two models are: (1) a time-indexed mixed-integer linear program, and (2) a novel binary optimization problem with linear and quadratic constraints and a linear objective. Our experiments indicate that optimization methods are very susceptible to modeling techniques and different solvers require dedicated methods. We show in this work that quantum-classical metaheuristics can benefit from a new way of modeling mathematical optimization problems. Additionally, we present a detailed performance comparison of the two solution methods for each optimization model. 
\end{abstract}


\section{Introduction}

The process industry is presently faced with uncertainty and pressures for product differentiation and accelerated product life cycles. In this context, modular production logistics emerges as a transformative approach to the planning and operation of production facilities within the industry. This paradigm shift enables companies to design their production processes with enhanced efficiency and flexibility, allowing them to quickly align production with the current demand dynamics.

Mathematical optimization is a key methodology for enhancing efficiency in the production process. In this paper, we present a job-shop scheduling problem arising from the project \textit{Modular Production Logistics} \cite{MoProLog}. The job-shop scheduling problem consists of two types of jobs, each with its own set of tasks.  Notably, one specific task from each job type is performed on the same machine. Additionally, the transportation of the jobs to and from the machines is carried out by autonomously guided vehicles (AGVs), which must be considered in the optimization.

In this work, we present two mathematical programming formulations for this problem. A more traditional mixed-integer programming model and a novel binary optimization problem with linear and quadratic constraints and a linear objective, where the linear and quadratic constraints can be easily moved to the objective, resulting in a quadric unconstrained binary optimization problem (QUBO). In addition, we assess and compare the performance of two solution methods, a quantum-classical optimization method (a metaheuristic), and a traditional mixed-integer programming solver for two mathematical programming formulations. 

Similar work has been done recently in \cite{awasthi2024realworldapplicationquantumclassical}. In this article, the authors introduced two different formulations for an optimization problem and compared a quantum-classical optimization method with a mixed-integer programming solver for an industry-relevant problem. The optimization problem is a scheduling problem in the chemical industry with the goal of uniformly using machines, minimizing setup times, and maximizing the value of processed jobs. 

In \cite{leibl}, the authors present a comprehensive case study that compares the performance of quantum-classical optimization methods, quantum-inspired annealing methods, and state-of-the-art classical mixed-integer programming solvers for a transport robot scheduling problem. As in this work, they also use different formulations for each of the solution methods.

Similar benchmarks for industry-relevant problems are published in~\cite{comp_cqm_cplex, portfolio_dwave_hybrid,Qgrid_demand_side,SCHWORM202313}. In \cite{osaba2024hybridquantumsolversproduction}, the authors compared different quantum-classical optimization solvers on four different combinatorial optimization problems. 

In the review article \cite{Yarkoni_2022}, many industry-relevant optimization problems are solved with (pure) quantum annealing. In \cite{Schworm2023, Carugno2022, SCHWORM2024142}, versions of the job-shop scheduling problems are solved by means of (pure) quantum annealing.

Our work provides
\begin{itemize}
    \item[1.] a novel formulation for an industry-relevant job-shop scheduling problem that can be solved for good solutions with a quantum-classical metaheuristic, and can easily be modified to other job-shop scheduling problems,
    \item[2.] a comprehensive study of the solution quality and runtime for each solution method applied to the two mathematical programming formulations, and
    \item[3.] the insight that new solution methods, such as quantum-classical metaheuristics, can benefit from new formulations.
\end{itemize}

\section{Problem definition}

\subsection{Problem description}

The problem is a variant of the job-shop scheduling problem that integrates transportation with AGVs. There are two types of jobs (A and B), each with different tasks that must be scheduled on different machines. In addition to scheduling the tasks, we also need to schedule the AGVs responsible for transporting jobs to the corresponding machines.

Each A-job consists of two tasks. The first tasks of the A-jobs are executed sequentially on machine $0$ in a predefined order and are therefore not part of the optimization. The second task of each A-job must be processed on machine $1$. 

Each B-job consists of three tasks. The first and last task must be processed on machine $1$ and the second task on machine $2$. 

For each task, an AGV is required to deliver the job to a machine and then pick it up after the task is completed. The only exceptions is the first task of each A-job, which only requires pickup. In a transportation job, the AGV starts at a depot (or warehouse), travels to the machine, and then returns to the depot. During this round trip, either a job can be delivered to a machine or picked up from a machine. One exception from the route 'depot-machine-depot' exists. The AGV can transport a job from one machine to another machine.

\begin{figure*}[t]
 \centering
 \includegraphics[width=\linewidth]{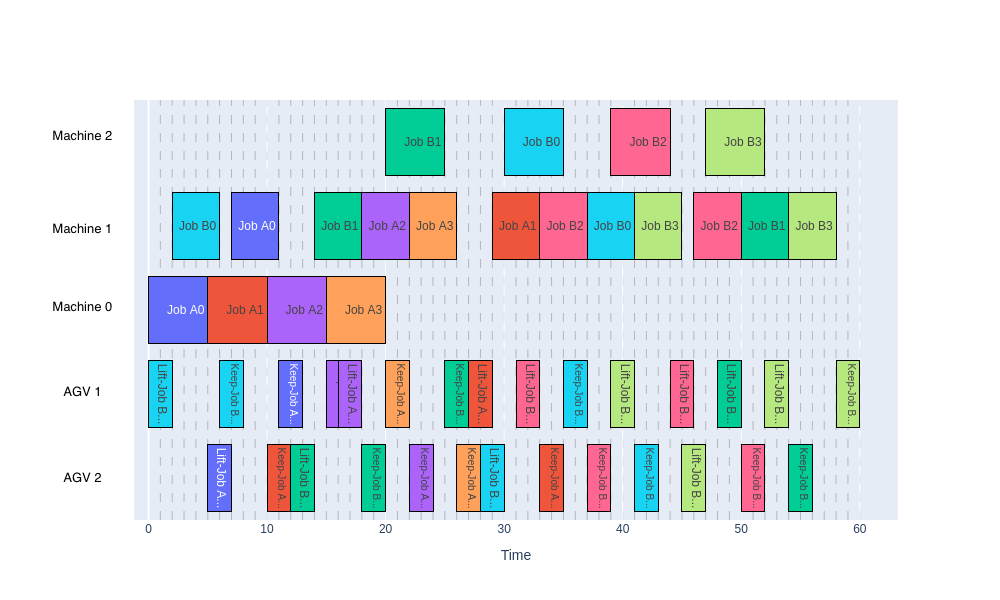}
 \caption{In the example, a feasible schedule is visualized. The instance has two AGVs, four A-jobs and four B-jobs. We can see that the first tasks of the A-jobs are scheduled consecutively on machine $0$. At $t=18$, job $B_1$ is brought from machine $1$ directly to machine $2$. The makespan of the solution is $60$.}
 \label{fig_problem}
\end{figure*}

The objective of the optimization problem is to assign feasible start times to each task and to allocate the corresponding transportation jobs to the AGVs, while minimizing the overall makespan. An example of a feasible schedule is illustrated in Figure \ref{fig_problem}.

The details of the optimization problem will be introduced when formulating the optimization problem as a mixed-integer linear program. Afterwards, a formulation with linear objective, and linear and quadratic constraints is introduced.

\subsection{Mixed-integer linear programming formulation}

In this section, we formulate the job-shop scheduling problem as a timed-index mixed-integer linear program (MILP) \cite{KU2016165}. Before we do this, we introduce notations. The tasks are separated by type: $A_k$ and $B_k$ are the sets of $k$-th tasks of A respectively B-jobs. We will mainly distinguish between $A_1$ tasks and all others:
\begin{align*}
    J &= A_2 \cup B_1 \cup B_2 \cup B_3, \\
    J_1 &= A_2 \cup B_1 \cup B_3, \\
    J_2 &= B_2.
\end{align*}
The set $J_m$ is the set of all tasks that must be processed on machine $m=1,2$.

To formulate the predecessor relation, we introduce the map $\sigma$ that maps a task to the corresponding job. That is $j \in B_k$ is the $k$-th task of the B-job $\sigma(j)$ and two tasks $j_1\in A_1,j_2\in A_2$ belong to the same job if $\sigma(j_1) = \sigma(j_2)$. Now, the set of all predecessor-successor relations is
$$ R = R_A \cup R_B $$
with 
$$ R_A = \{ (j_1,j_2) \in A_1 \times A_2 : \sigma(j_1) = \sigma(j_2) \} $$
being the predecessor-successor of the A-jobs and 
$$ R_B = \{ (j_1,j_2) \in B_{12} \cup B_{23} : \sigma(j_1) = \sigma(j_2) \} $$ 
the predecessor-successor of the B-jobs, with $B_{12}=B_1 \times B_2$ and $B_{23} = B_2 \times B_3$.

The set of all AGVs is $K$ and each AGV trip needs $\delta \in \mathbb{N}$ time steps. Meaning, an AGV needs $\delta$ time steps to go from the depot to a machine, or return from a machine to the depot, or go from a machine to another machine.

Furthermore, $T=\{1,\ldots,|T|\}$ is the set of all time steps with $|T|$ being the time horizon. 

The first task of each A-job ends in $\tau_j, j \in A_1$, that is, $\tau_j$ is the smallest time step at which it is no longer processed. For example, a task that starts in $t-p$, with $p$ being the processing time, ends in $t$ and is busy in the time steps $t-p,\ldots,t-1$.

\subsubsection{Variables}
To model the problem we need three types of variables. The variables 
$$x \in \{0,1\}^{T \times J \times K \times K}$$ 
model the start and transportation of all tasks and 
$$x_{t,j,k_1,k_2} = 1$$
if and only if task $j \in J$ starts at time $t \in T$ and is delivered to the machine by AGV $k_1 \in K$ and picked up by AGV $k_2\in K$. 

Since the start of any task $j\in A_1$ is fixed, it is only necessary to decide the AGV used for pick up:
$$y \in \{0,1\}^{A_1 \times K},$$
with 
$$y_{j,k}=1$$ 
if and only if task $j\in A_1$ is picked up by AGV $k\in K$. 

The makespan is modeled by the (positive) real variable
$$c \geq 0 .$$

\subsubsection{Constraints}
In a feasible solution, all tasks must be scheduled exactly once, hence,
$$ \sum_{t \in T} \sum_{k_1, k_2 \in K}x_{t,j,k_1,k_2} = 1 $$
for all tasks $j\in J$ and 
$$ \sum_{k \in K} y_{j,k} = 1$$
for all tasks $j \in A_1$. 

Furthermore, at each time $t\in T$ only one task can be processed per machine $m \in \{1,2\}$:
$$\sum_{j \in J_m} \sum_{s\in (t-p_j+1)^{++}}^t \sum_{k_1, k_2 \in K} x_{s,j,k_1,k_2} \leq 1.$$
For convenience we use $x^{++} = \max\{x,1\}$.

It also must be ensured that a successor can only start when its predecessor finished:
\begin{align*}
\sum_{t \in T} \sum_{k_1, k_2 \in K} (t+p_i+\delta)x_{t,i,k_1,k_2} \\ 
\leq \sum_{t \in T} \sum_{k_1, k_2 \in K} t x_{t,j,k_1,k_2}
\end{align*}
for any predecessor-successor pair $(i,j)\in R$.

$$ \sum_{j \in J} \sum_{k_2 \in K} x_{t,j,k_1,k_2} \leq 1.$$

If a task starts or ends at $t \in T$ and the transportation job is carried out by AGV $k \in K$, no other task can use AGV $k$ between and including $t-\delta+1$ and $t+\delta$. That means only one task can start within the mentioned time interval:
$$ \sum_{s = (t-\delta+1)^{++}}^{t+\delta} \sum_{j \in J} \sum_{k_2 \in K} x_{s,j,k,k_2} \leq 1 $$
and only one task can end within this interval:
\begin{align*}
\sum_{j \in J} \sum_{s=(t-p_j-\delta+1)^{++}}^{(t-p_j+\delta)^{++}} \sum_{k_1 \in K} x_{s,j,k_1,k} \\ 
+ \sum_{\substack{j \in A_1 \\ t-\delta+1 \leq \tau_j \leq t+\delta }} y_{j,k} \leq 1 .
\end{align*}

If a task $j\in J$ starts in $t\in T$, a second task $i \in J \setminus \{j\}$ cannot end between and including $t-2\delta+1$ and $t$, i.e., not start in $\{t-p_i-2\delta+1,\ldots,t-p_i\}$, if it is not the direct predecessor and uses the same AGV:


\begin{align*}
\sum_{\substack{i \in A_1: (i,j) \not\in R\\ \mathclap{t-2\delta+1 \leq \tau_i\leq t}}} y_{i,k} + \sum_{i \in J: (i,j) \not\in R} \sum_{s=(t-p_i-2\delta+1)^{++}}^{t-p_i} x_{s,i,k_1,k} \\
+ \sum_{k_2 \in K} x_{t,j,k,k_2} \leq 1.
\end{align*}

The first line counts the tasks that end in $\{t-2\delta+1,\ldots,t\}$ and use AGV $k$. The left-hand-side in the second line counts the tasks that start in $t$ and use AGV $k$.  

In addition, an AGV cannot deliver a job to a machine and then pick up another job from the same machine during a single run. More precise, if a task starts at $t \in T$ and uses AGV $k \in K$, another task cannot finish right before it in $t-1$ and be picked up by the same AGV $k$:
\begin{align*}
\sum_{\substack{ j \in J \\  t-p_j \geq 1}} \sum_{k_1 \in K} x_{t-p_j,j,k_1,k} \\ 
+ \sum_{j \in J} \sum_{k_2 \in K} x_{t,j,k,k_2}  \leq 1.
\end{align*}

Last but not least, the last finished task, respectively the transportation afterwards, defines the makespan:
$$\sum_{t \in T} \sum_{k_1, k_2 \in K}  (t+p_j) x_{t,j,k_1,k_2} + \delta \leq c $$
for all jobs $j \in J$ (that do not have a successor).

\subsubsection{Objectives} The objective is to minimize the makespan by minimizing $c$. 

\subsection{Quadratic constrained binary optimization formulation}

In this section, we introduce a novel quadratic constrained binary optimization (QCBO) formulation.  The problem is formulated as a pure binary program with a linear objective, and linear and quadratic constraints.

\subsubsection{Variables} 

For each task, a variable for the start and end of the task, and the AGV that is used for the transportation is introduced. The variables
$$ x \in \{0,1\}^{J \times T \times K}$$
model the start of each task and the AGV used to transport the job to the machine with 
$$x_{j,t,k}=1$$
if and only if task $j \in J$ starts in $t \in T$ and AGV $k \in K$ is used for the transportation. Similar,   
$$ y \in \{0,1\}^{J \times T \times K}$$
models the end of the tasks with 
$$y_{j,t,k}=1$$
if and only if the task $j \in J$ ends in $t \in T$ and AGV $k \in K$ is used for the transportation. 

For all tasks $j \in A_1$ we introduce the variables 
$$ z \in \{0,1\}^{A_1 \times K}$$
with 
$$z_{j,k}=1$$ 
if task $j \in A_1$ is picked up by AGV $k\in K$.

The makespan is modeled by
$$u \in \{0,1\}^T$$
and $u_t = 1$ if the last job ends in $t$.

\subsubsection{Linear constraints} 

Every task $j\in A_1$ must be picked up: 
$$ \sum_{k \in K} z_{j,k} = 1.$$
Similar, every task $j \in J$ must start and end. Hence, 
$$ \sum_{t \in T} \sum_{k \in K} x_{t,j,k} = 1$$
and 
$$ \sum_{t \in T} \sum_{k \in K} y_{t,j,k} = 1 .$$
If task $j \in J$ starts in $t =1,\ldots,|T|-p_j$, it must end in $t+p_j$: 
$$ \sum_{k \in K} x_{t,j,k} = \sum_{k \in K} y_{t+p_j,j,k}.$$

\subsubsection{Quadratic constraints} 

To ensure that the predecessor-successors relations and the minimum traveling times are respected, a task cannot end after $t-\delta+1$ if the successor starts in $t \in T$:
$$ \left( \sum_{k \in K} x_{t,j,k}  \right) \left( \sum_{s = t-\delta+1}^{|T|} \sum_{k \in K} y_{s,i,k}  \right) = 0 $$
for all $(i,j) \in R$.

Two tasks that are processed on the same machine cannot be processed at the same time. For each machine $m \in \{1,2\}$ and job $j \in J_m$ starting at $t =1, \ldots, |T|$ no other job $i\in J_m \setminus \{1,2\}$ can start in $\{t, \ldots, t+p_j-1\}$:
$$ \left( \sum_{k \in K} x_{t,j,k} \right) \left( \sum_{s=t}^{t+p_j-1 }\sum_{i \in J_m \setminus \{j\}}\sum_{k \in K} x_{s,i,k} \right) = 0 .$$

To respect the travel times of an AGV, a minimum margin of $2\delta$ between two tasks must be maintained. To model this requirement, we consider four cases: the transportation takes place (1) between the start times of two jobs, (2) between the end times of two jobs, (3) between the start of one job and the end of another, and (4) between the end of one job and the start of another. For case (4), if the two jobs are direct predecessors, only a margin of $\delta$ must be maintained.

We consider the AGV $k\in K$. If a task starts at time $t \in T$, no other task can start between and including $t$ and $t+2\delta-1$. As a quadratic constraint:
$$  \left( \sum_{j \in J} x_{t,j,k} \right) \left( \sum_{s=t}^{t+2\delta-1} \sum_{i \in J \setminus \{j\}}  x_{s,i,k} \right) = 0.$$
Similarly, if task ends at $t \in T$, no other task can end between and including $t$ and $t+2\delta-1$:
\begin{align*}
 &\left( \sum_{j \in J} y_{t,j,k} + \sum_{j \in A_1: \tau_j=t} z_{j,k}\right)  \\ 
\cdot &\left( \sum_{s=t}^{t+2\delta-1} \left( \sum_{i \in J \setminus \{j\}} y_{s,i,k} + \sum_{i \in A_1: \tau_j=s} z_{i,k}\right) \right) = 0.
\end{align*}
Furthermore, if a task starts at time $t \in T$, no other task can end between and including $t$ and $t+2\delta-1$:
\begin{align*}
 \left( \sum_{j \in J} x_{t,j,k} \right) \left( \sum_{s=t}^{t+2\delta-1} \left( \sum_{i \in J \setminus \{j\}} y_{s,i,k} + \sum_{i \in A_1: \tau_j=s} z_{i,k}\right) \right)
 \\= 0.
\end{align*}
For the last case, if a task ends at time $t \in T$, no other task can start between and including $t$ and $t+2\delta-1$ but the direct successor:
$$\left( \sum_{j \in J} y_{t,j,k} + \sum_{j \in A_1: \tau_j=t} z_{j,k}\right)   
\left( \sum_{s=t}^{t+2\delta-1}  \sum_{\substack{i \in J \setminus \{j\}  \\ (i,j) \not\in R}} x_{s,i,k} \right) = 0.$$
For each pair $(j,i)\in R$, the equation reads:
$$ y_{t,j,k} \left(\sum_{s=t}^{t+\delta-1}x_{s,i,k}\right) = 0$$
if $j\in J$ and 
$$ z_{j,k} \left(\sum_{s=t}^{t+\delta-1}x_{s,i,k}\right) = 0$$
if $j \in A_1$ and $\tau_j=t$.

For a predecessor-successor pair $(i,j)\in R$, the minimum margin between $i$ and $j$ is $2\delta$ if different AGVs are used for pick up and delivery. Hence, if $i$ is picked up by AGV $k \in K$,
$$ y_{t,i,k} \left(\sum_{s=t}^{t+2\delta-1}\sum_{l\in K\setminus \{k\}} x_{s,j,l}\right) = 0$$
if $j\in J$ and 
$$ z_{i,k} \left(\sum_{s=t}^{t+2\delta-1}\sum_{l\in K\setminus \{k\}}x_{s,i,l}\right) = 0$$
if $j \in A_1$ and $\tau_j=t$.

For the makespan variables $u$, 
$$ \sum_{t \in T} u_t = 1$$
must hold. They are coupled with the $y$ variables as follows. If a task $j\in B_3 \cup A_2$ ends in $t$, the makespan cannot be smaller than $t$, i.e., $u_s = 0$ for all $s \in T, s < t$. This translates to the following quadratic constraint:
$$\left( \sum_{k \in K} y_{t,j,k} \right)  \left( \sum_{s=1}^{t-1} u_s \right) = 0  $$
for any task $j\in B_3 \cup A_2$.

Notice that the left-hand sides of all quadratic constraints are nonnegative. Hence, the equality constraints ($=0$) can be changed to inequality constraints ($\leq 0$). Also, the quadratic constraints can be aggregated (further) and together with the linear constraint and a big enough penalty can be moved to the objective to obtain a QUBO problem.

\subsubsection{Objective} To minimize the makespan,
$$\sum_{t \in T} t u_t + \delta$$
must be minimized.

\section{Experiments}

The purpose of the experiments is to evaluate how different formulations of our optimization problem affect the performance of the solution process. Specifically, we compare the performance of the D-Wave solver on each model and compare it with the performance of Gurobi, a state-of-the-art classical solver.

\subsection{Optimization solvers}

\subsubsection{Gurobi}
Gurobi \cite{gurobi} is a powerful mathematical programming solver that utilizes the branch-and-bound algorithm; see, for example, \cite{integer_programming}. It is considered state-of-the-art and is designed to run on classical hardware. This algorithm works by breaking down the original problem into smaller subproblems using a decision tree. By using lower bounds derived from linear programming relaxations, Gurobi is able to reduce the search space and improve efficiency. Gurobi is an all-purpose solver, which means it can be used to solve a wide range of optimization problems. It is capable of finding a guaranteed optimal solution in a deterministic manner, given enough time. 

In this work, we used the latest Gurobi version 12.0.1 on a MacBook Pro with an Apple M1 Max CPU with a clock speed of 3.20 GHz featuring eight performance cores and two energy-efficient cores, and 32 GB of memory.\footnote{We also used a Gurobi provided cloud service that utilizes Gurobi 11.0.0. to run the experiments. There, an Intel(R) Xeon(R) Platinum 8124M CPU with a clock speed of 3.00 GHz and 8 physical cores was used. In our experiments, in $10\%$ of the runs the cloud service outperformed the MacBook, in $52\%$ of the runs the MacBook outperformed the cloud service, and the remaining runs yield the same solution quality. }

\subsubsection{D-Wave Leap’s Hybrid solver}

D-Wave provides quantum and hybrid solvers, which combines quantum and classical computing, to solve optimization problems \cite{dwave_solver} and several quantum annealers via cloud-based access, which have over $5000$ qubits and an average of $15$ couplers per qubit \cite{farre}. These quantum annealers are designed to solve QUBO problems. In a QUBO problem, each optimization variable is represented by a qubit, and the optimal solution corresponds to the ground state of the system. The quantum annealing mechanism employed by D-Wave aims to find the ground state by evolving the quantum system over time and then measuring all the qubits to reveal the optimal solution. 

However, D-Wave's quantum hardware has limited connectivity, meaning that each qubit can only interact with a certain number of other qubits. This limitation imposes constraints on the correlations that can be represented by the Hamiltonian encoding the QUBO problem. Additionally, finding a suitable representation of the problem with these connectivity constraints is a computationally challenging task known as an NP-hard problem. To overcome this challenge, D-Wave's CQM solver utilized both classical and quantum hardware. In this context, CQM stands for Constrained Quadratic Model, which can handle both quadratic objectives and quadratic constraints.

The CQM solver allows for the solution of larger optimization problems that exceed the capacity of the quantum hardware by decomposing the original problem into smaller parts and presenting only a subset of the problem to the quantum annealer. However, the exact workings of CQM are not publicly disclosed.

\subsection{Instances and Runtime}

\subsubsection{Instances}
For benchmarking, we created different instances by varying the number of A- and B-jobs, the number of available AGVs, the processing times of the tasks, and the time horizon. The ranges of parameters are as follows:
\begin{itemize}
    \item The number of $A$ and $B$ jobs ranges from $3$ to $11$ with maximum number of total jobs being $22$.
    \item Up to $5$ AGVs are considered.
    \item The time horizon is at most $541$ and large enough to allow a feasible solution in all instances.
    \item The processing time of the first $A$ job is between $3$ and $9$ and for the second $A$ job between $2$ and $8$.
    \item The processing time of the first $B$ job is between $2$ and $9$, for the second $B$ job between $3$ and $8$, and for the last $B$ job between $2$ and $7$.
\end{itemize}

\subsubsection{Runtimes} 
D-Wave’s metaheuristic does not include an automatic stopping criterion and therefore does not stop on its own; it requires an explicitly defined runtime. Moreover, for each instance, D-Wave can compute an instance-specific minimum runtime that depends on the size of the problem \cite{dwave_timer}. For D-Wave’s metaheuristic, we will use the minimum runtime, as well as two times and five times that runtime, with each rounded up to the next second\footnote{In the following, 1x, 2x, and 5x correspond to one, two, and five times the from D-Wave computed instance-specific runtime. When we say, for example, 2x the runtime, we mean two times the standard runtime.}. The instance-specific runtime is computed with respect to the QCBO model because it is greater than the minimum instance-specific runtime of the MILP model. Since D-Wave has a minimum runtime of five seconds, we only considered instances with a minimum runtime of (strictly) more than five seconds. We also use this runtime as a proxy of the size of the problem and our instances are ordered in increasing size. 

\subsection{Performance measures}

\subsubsection{Solution quality}

The performance measure \textit{solution quality} measures the quality of the solution computed after a certain runtime. In our experiment the runtimes for each instance are the instance-specific runtime from D-Wave, twice and five times this runtime.

We measure the quality of the solution through the primal gap and wins. Given the best known objective value $\gamma^* \in \mathbb{N}$, the \textit{primal gap} for the objective value $\gamma \geq \gamma^*$ is
$$ \text{gap}(\gamma) = 100 \cdot \left( \frac{\gamma -\gamma^*}{\gamma^*} \right) [\%] .$$
The (primal) gap is always positive and a gap of $\text{gap}(\gamma)=10$ means that $\gamma$ is $10\%$ worse than the best known solution $\gamma^*$. The best known solution is the best known solution we computed during the considered experiments. 

In addition to the gap, we consider wins. For two solution methods $A$ and $B$, let $\lambda_A$ and $\lambda_B$ be the objectives (makespan). We say method $A$ wins if $\lambda_A < \lambda_B$. If $\lambda_A = \lambda_B$, the two methods tie.

By convention, if a solver could not find a feasible solution, we set the objective to the makespan of the trivial solution, which is scheduling all jobs sequentially, which is the sum of all processing and transportation times. For A-jobs three transportation jobs and for B-jobs six transportation jobs are needed. Therefore, the makespan is given by
$$ \sum_{j\in J \cup J_A} p_j + n_A \cdot (3\delta) + n_B \cdot (6\delta) .$$

\subsubsection{Time to solution}

The \textit{time to solution} refers to the time it takes for an algorithm to solve a given problem to a certain solution quality. In our experiments, the solution quality is given by the best solution computed by D-Wave. Two compare the time to solution we compute the quotient 
$$ \Delta(t_{\text{D-Wave}},t_{\text{Gurobi}}) = \frac{t_{\text{D-Wave}}}{t_{\text{Gurobi}}},$$
with $t_{\text{D-Wave}}$ being the time to solution for D-Wave and $t_{\text{Gurobi}}$ for Gurobi. If $\Delta(t_{\text{D-Wave}},t_{\text{Gurobi}}) > 1$, D-Wave was slower by a factor of $\Delta(t_{\text{D-Wave}},t_{\text{Gurobi}})$.  

\subsubsection{Number of runs} To obtain reliable performance statistics, we repeat each optimization five times. For D-Wave’s inherently random method, all five runs use the same parameters. For Gurobi, we instead vary the random‐number seed across the five runs.

\subsection{Comparing the performance of D-Wave on the QCBO and MILP model}

\begin{figure*}[hbtp]
    \centering
    \subfigure{\includegraphics[width=0.49\textwidth]{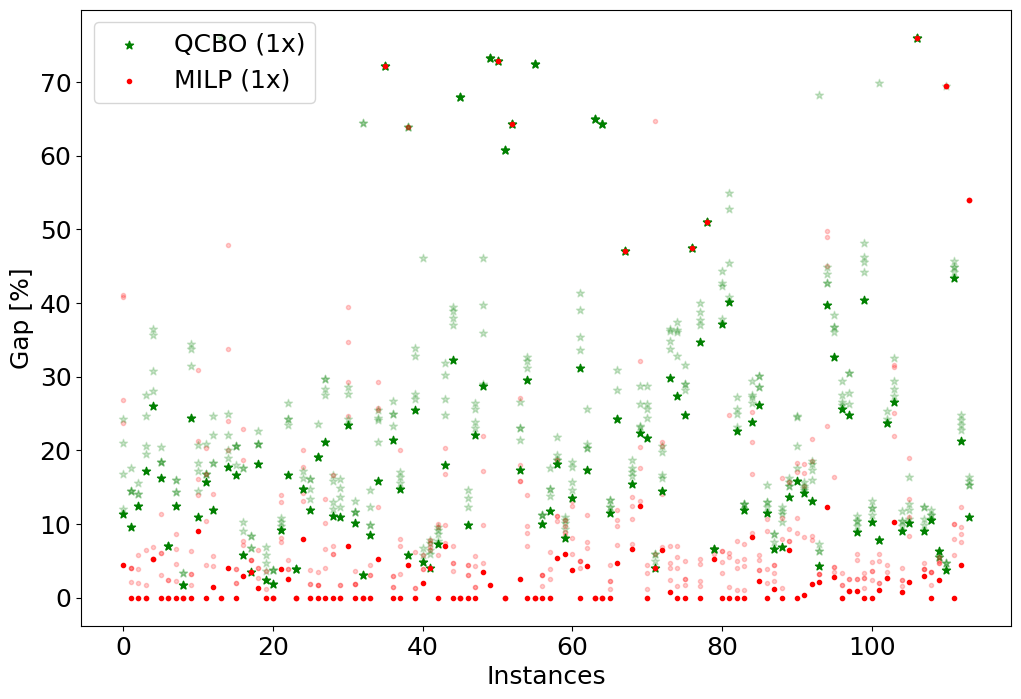}}  
    \subfigure{\includegraphics[width=0.49\textwidth]{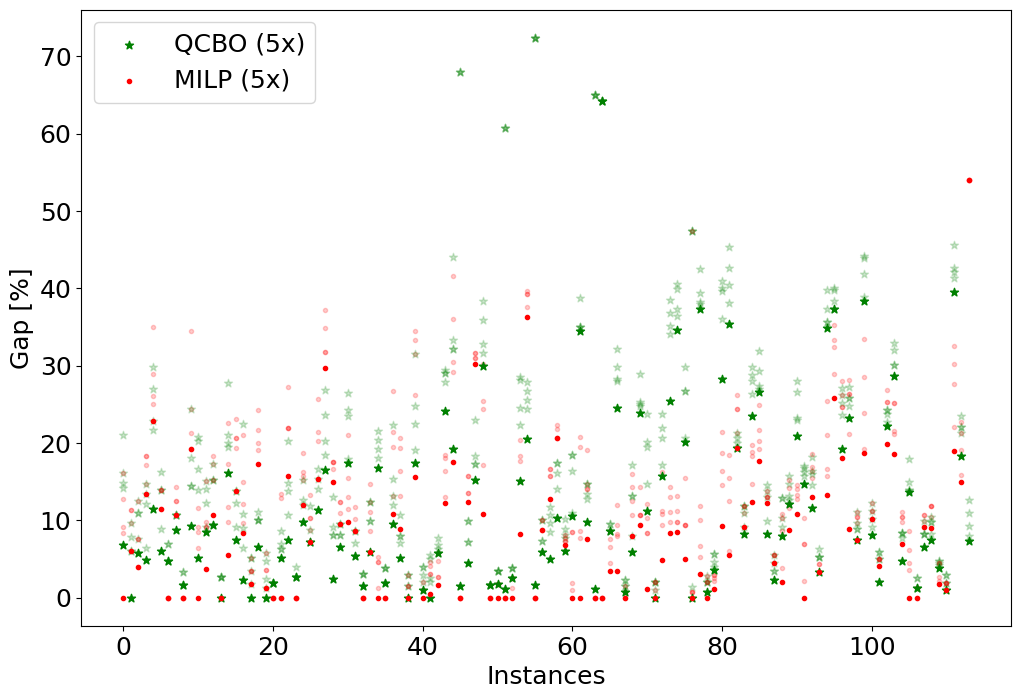}}
    \caption{The two figures compare the solution quality of D-Wave on the two different formulations, the QCBO and the MILP formulations, for every instance. The left for the 1x runtimes, the right for the 5x runtimes. The best runs are indicated by the dark stars and the dark dots, respectively.}
    \label{fig:dwave_gap}
\end{figure*}

First, we evaluate the performance of D-Wave on the two formulations. Because we ran D-Wave  for a specific runtime, we can only compare the solution quality and not the time to solution. In Table \ref{table_gap_dwave_statistics}, the statistical quantities are shown and in Figure \ref{fig:dwave_gap} the gaps are visualized for the 1x and 5x runtimes.

\begin{table}[ht]
\begin{center}
{\small
\begin{tabular}{ |c|c|c|c|c| } 
 \hline
 &  \multicolumn{2}{c|}{MILP} & \multicolumn{2}{c|}{QCBO} \\
 \hline
 &  Average & \thead{Standard \\ deviation} & Average & \thead{Standard \\ deviation} \\ 
  \hline
 1x (best) & $6.71$ & $16.59$ & $22.24$ & $18.71$ \\ 
 2x (best) & $3.60$ & $9.65$ & $15.82$ & $15.62$\\  
 5x (best) & $8.29$ & $8.77$ & $11.58$ & $11.42$\\ 
  \hline
 1x (median) & $12.61$ & $18.88$ & $25.09$ & $18.73$ \\ 
 2x (median) & $9.37$ & $14.71$ & $20.79$ & $17.61$ \\ 
 5x (median) & $11.79$ & $9.85$ & $17.47$ & $15.26$ \\  
  \hline
 1x (worst) & $17.17$ & $19.78$ & $29.79$ & $19.47$ \\ 
 2x (worst) & $13.46$ & $16.15$ & $25.50$ & $19.01$ \\  
 5x (worst) & $15.14$ & $11.40$ & $20.09$ & $15.58$ \\ 
 \hline
\end{tabular}
}
\end{center}
\caption{The table compares the average gap and the standard deviation for D-Wave solving the QCBO and MILP model, for the best, median, and worst runs.}
\label{table_gap_dwave_statistics}
\end{table}

The figure clearly shows that, at the standard runtime (1x), D-Wave performs better on the MILP model than on the QCBO model. This is not as evident for the 5x runtimes, but still present. These observations are supported by the statistical quantities, which also show that the MILP model still outperforming the QCBO when comparing the average gap. The narrowing of the gaps with increased runtime is not solely due to the expected performance improvements on the QCBO model. It also reflects the fact that the average gap for the MILP model does not necessarily decrease with longer runtime. For the best runs, the average gap even increases. Looking at the statistical quantities, we can see that the average gap for the MILP model is the smallest for the MILP model run with the 2x runtimes, and not with the 5x runtimes, which would be the expected behavior.

As noted earlier, for the 5x runtimes, the performance difference between the MILP and QCBO models is not as clear as for the 1x runtimes. When counting wins, the MILP model wins in $56.67\%$ of the runs and the QCBO model wins in $34.39\%$ of the runs. The remaining runs end in a draw. We always compared the best runs, the second-best run, etc. Analyzing the instances, we could not find a dependency between the model parameters and the relative performance of the MILP model and the QCBO model.

\begin{figure}[ht]
    \centering
    \subfigure{\includegraphics[width=0.8\textwidth]{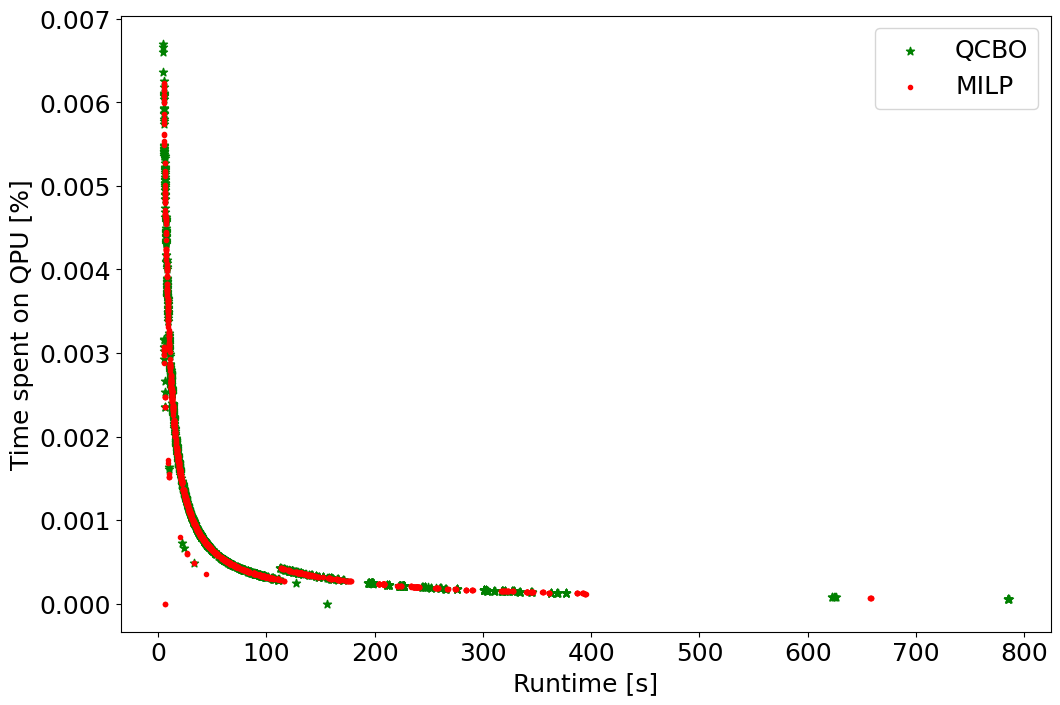}}
    \caption{The figure shows the relative time spent on the QPU for each run for the MILP and QCBO model.}
    \label{fig:dwave_qpu_time}
\end{figure}

D-Wave is a quantum-classical metaheuristic that utilizes quantum computing, in this case quantum annealing, to enhance the performance of the metaheuristic. Due to D-Wave being proprietary, we do not know the influence of quantum computing on the solution process. However, D-Wave provides the time spent on the QPU, which can be seen in Figure \ref{fig:dwave_qpu_time} and shows that the time spent in the QPU is negligible compared to the total runtime and the relative time spent on the QPU decreases for increasing runtime.

\subsection{Comparing the performance for the MILP}

We now compare the performance of D-Wave with the performance of Gurobi on the MILP model. We will see that Gurobi outperforms D-Wave, especially for increased runtimes.

\begin{figure*}[hbtp]
    \centering
    \subfigure{\includegraphics[width=0.49\textwidth]{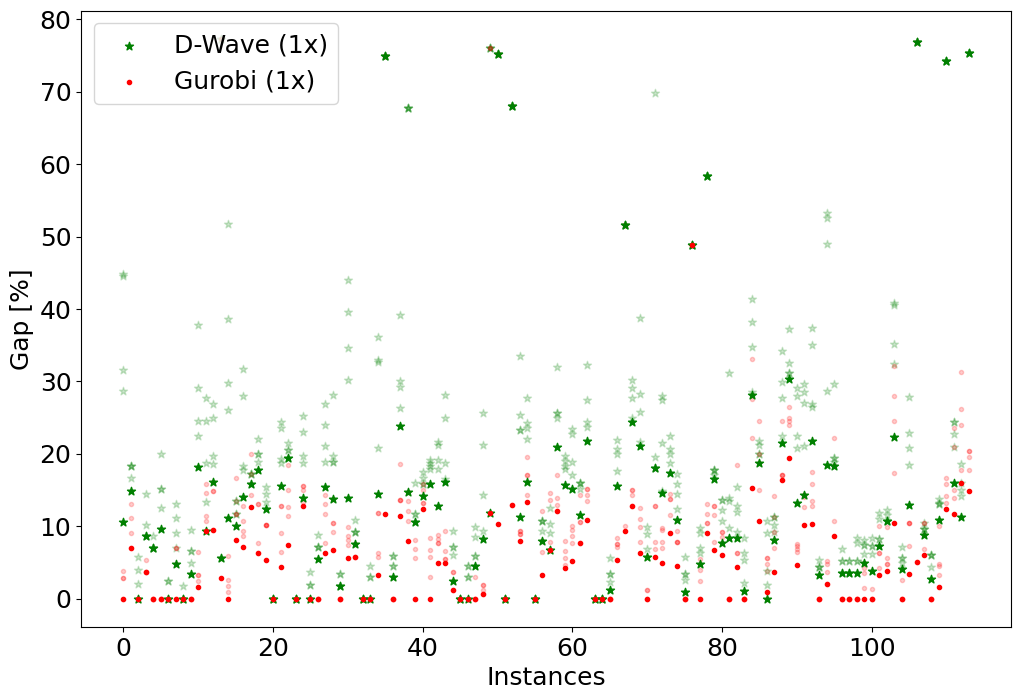}} 
    \subfigure{\includegraphics[width=0.49\textwidth]{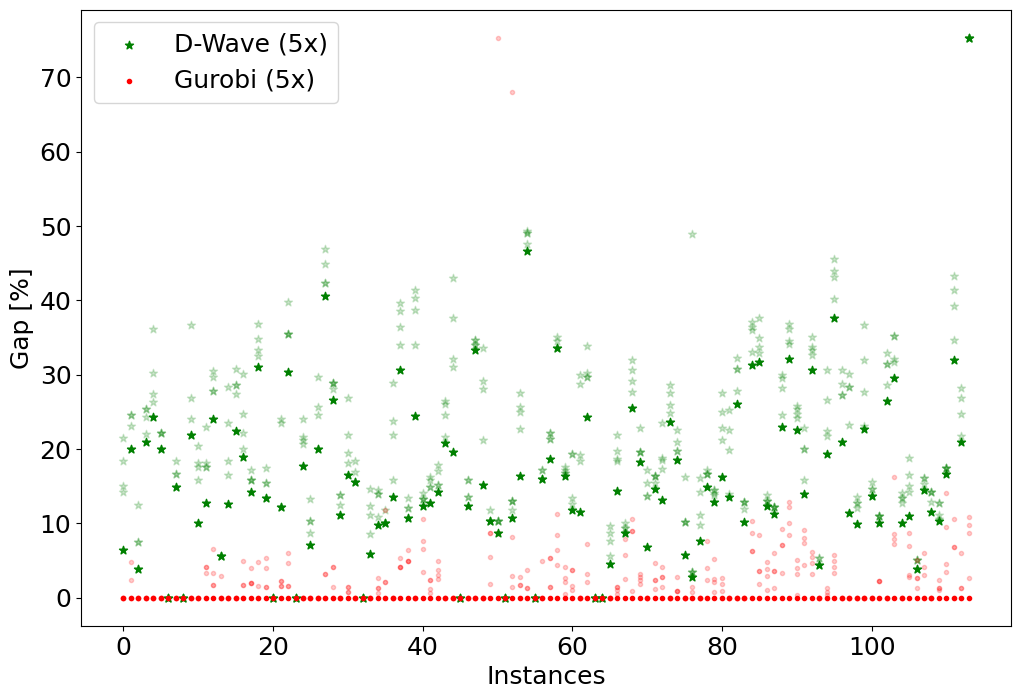}}
    \caption{The two figures compare the solution quality (gaps) of D-Wave and Gurobi for the MILP formulation for every instance. The left for the 1x runtimes, the right for the 5x runtimes. The best runs are indicated by the dark stars and the dark dots, respectively.}
    \label{fig:milp_gap}
    
\end{figure*}

First, we compare the solution quality. In Figure \ref{fig:milp_gap}, the gaps are visualized for the 1x and 5x runtimes and in Table \ref{table_gap_milp_statistics} the statistical quantities are presented. 

Again, we see that for increasing runtime D-Wave performance gets worse, relatively. The difference between the gaps increases from $9.35$ percentage points to $16.19$ percentage points when comparing the best runs, and even more when comparing the other runs. Again, the reason here is that Gurobi continuously improves the objective with increasing runtime and D-Wave does not.

\begin{table}[hb]
\begin{center}
{\small
\begin{tabular}{ |c|c|c|c|c| } 
 \hline
 &  \multicolumn{2}{c|}{Gurobi} & \multicolumn{2}{c|}{D-Wave} \\
 \hline
 &  Average & \thead{Standard \\ deviation} & Average & \thead{Standard \\ deviation} \\ 
  \hline
 1x (best) & $5.37$ & $6.41$ & $14.72$ & $17.11$ \\ 
 2x (best) & $2.91$ & $5.45$ & $12.01$ & $11.20$\\  
 5x (best) & $0.00$ & $0.00$ & $16.19$ & $11.10$\\ 
  \hline
 1x (median) & $11.39$ & $17.23$ & $20.05$ & $19.51$ \\ 
 2x (median) & $5.38$ & $8.92$ & $17.24$ & $15.29$ \\ 
 5x (median) & $1.73$ & $2.43$ & $19.35$ & $11.94$ \\  
  \hline
 1x (worst) & $13.46$ & $17.33$ & $24.10$ & $20.33$ \\ 
 2x (worst) & $8.66$ & $13.48$ & $20.91$ & $16.69$ \\  
 5x (worst) & $4.46$ & $9.77$ & $22.42$ & $12.99$ \\ 
 \hline
\end{tabular}}
\end{center}
\caption{The table compares the average gap and the standard deviation for Gurobi and D-Wave solving the MILP model, for the best, median, and worst runs.}
\label{table_gap_milp_statistics}
\end{table}

When comparing wins (as in the last section), we can see that for 1x the runtimes, Gurobi wins in $85.29\%$ of the runs and D-Wave in just $0.33\%$, which corresponds to two runs. 

\begin{figure}[ht]
    \centering
    \subfigure{\includegraphics[width=0.99\textwidth]{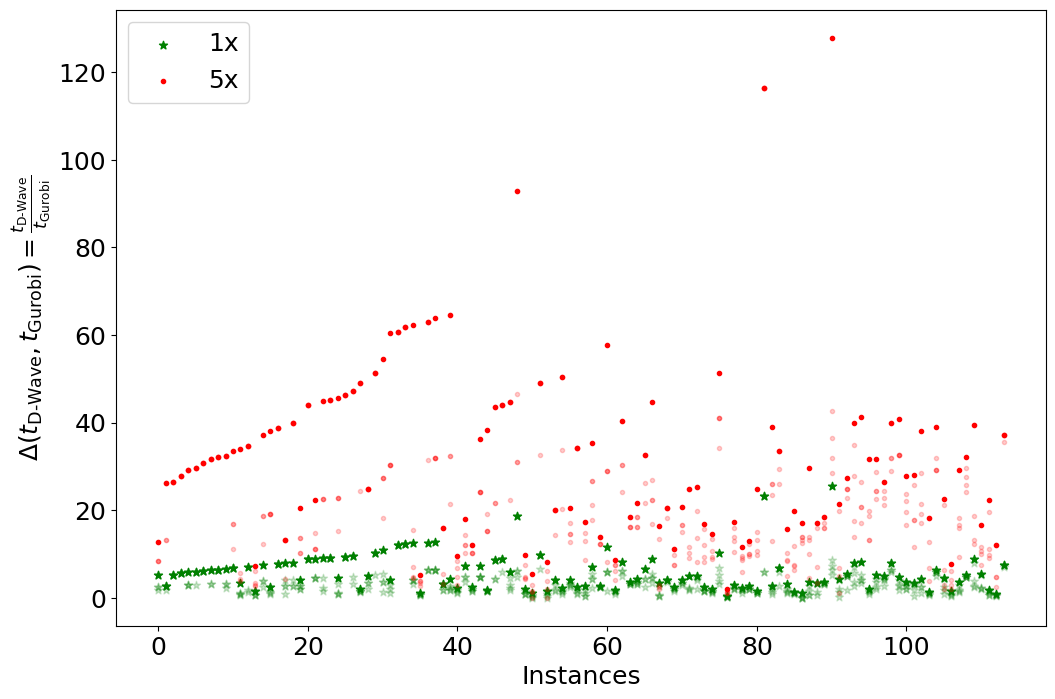}}
    \caption{The figure shows how much slower D-Wave is compared to Gurobi on the MILP model (if $\Delta(t_{\text{Gurobi}}, t_{\text{D-Wave}}) > 1$). For the first $40$ instances, the majority of the instances run $t_{\text{D-Wave}}$ faster on Gurobi.}
    \label{fig:milp_time}
\end{figure}

For the time to solution, the performance advantages of Gurobi are even more pronounced, especially for smaller instances. For these instances, Gurobi found the solution within the first second, making D-Wave $x$ times slower, with $x$ being the instance-specific runtime. From this follows directly, that for 5x runtimes, D-Wave is five times worse than for the standard runtimes. Table \ref{table_time_dwave_statistics} states the statistical values and runtime increases are visualized in Figure \ref{fig:milp_time}. For the statistical quantities, we chose the geometric average and standard deviation due to the following: Assume that for one run we have a runtime increase of $4$ and for another run we have an increase of $0.25$, which is a slowdown by $4$. Ideally, these effects should cancel each other out, which is not the case for the arithmetic mean, but it is for the geometric mean. The increasing geometric mean for $\Delta(t_{\text{Gurobi}}, t_{\text{D-Wave}})$ is due to the fact that for increasing runtimes, D-Wave does not find significant better solutions, i.e., the solution quality stagnates. However, the increase in runtime affects only the dominator in the expression $\Delta(t_{\text{Gurobi}}, t_{\text{D-Wave}}) = t_{\text{Gurobi}}/t_{\text{D-Wave}}$

\begin{table}[ht]
\begin{center}
{\small
\begin{tabular}{ |c|c|c| } 
 \hline
 &  Average & \thead{Standard \\ deviation} \\ 
  \hline
 1x (best) & $4.50$ & $2.08$ \\ 
 2x (best) & $8.86$ & $2.10$ \\  
 5x (best) & $27.04$ & $1.92$ \\ 
  \hline
 1x (median) & $2.89$ & $2.59$ \\ 
 2x (median) & $5.83$ & $2.35$ \\ 
 5x (median) & $18.63$ & $2.21$ \\  
  \hline
 1x (worst) & $1.73$ & $3.52$ \\ 
 2x (worst) & $3.44$ & $3.55$ \\  
 5x (worst) & $12.59$ & $3.41$ \\ 
 \hline
\end{tabular}
}
\end{center}
\caption{The table shows the geometric average and geometric standard deviation of runtime increases for D-Wave over Gurobi on the MILP model.}
\label{table_time_dwave_statistics}
\end{table}

We only compared the time to solution to best D-Wave run. When only comparing the best runs, D-Wave outperformed Gurobi in two runs. The remaining data can be found in Table \ref{table_time_dwave_wins}.

\begin{table}[ht]
\begin{center}
{\small
\begin{tabular}{ |c|c|c|c|c|c| } 
 \hline
 run &  best & \thead{second \\ best} & median & \thead{second \\ worst} & worst \\ 
  \hline
 1x & $2$ & $5$ & $12$ & $16$ & $29$\\ 
 2x & $1$ & $1$ & $3$ & $5$ & $11$\\  
 5x & $0$ & $0$ & $0$ & $0$ & $2$\\ 
 \hline
\end{tabular}
}
\end{center}
\caption{The table shows how often the best D-Wave run was faster then the Gurobi runs on the MILP model. For example, for 2x the runtime, the best D-Wave run was faster (just) $11$ times than the slowest (worst) Gurobi run.}
\label{table_time_dwave_wins}
\end{table}

\subsection{Comparing the performance for the QCBO}

\begin{figure*}[hbtp]
    \centering
    \subfigure{\includegraphics[width=0.49\textwidth]{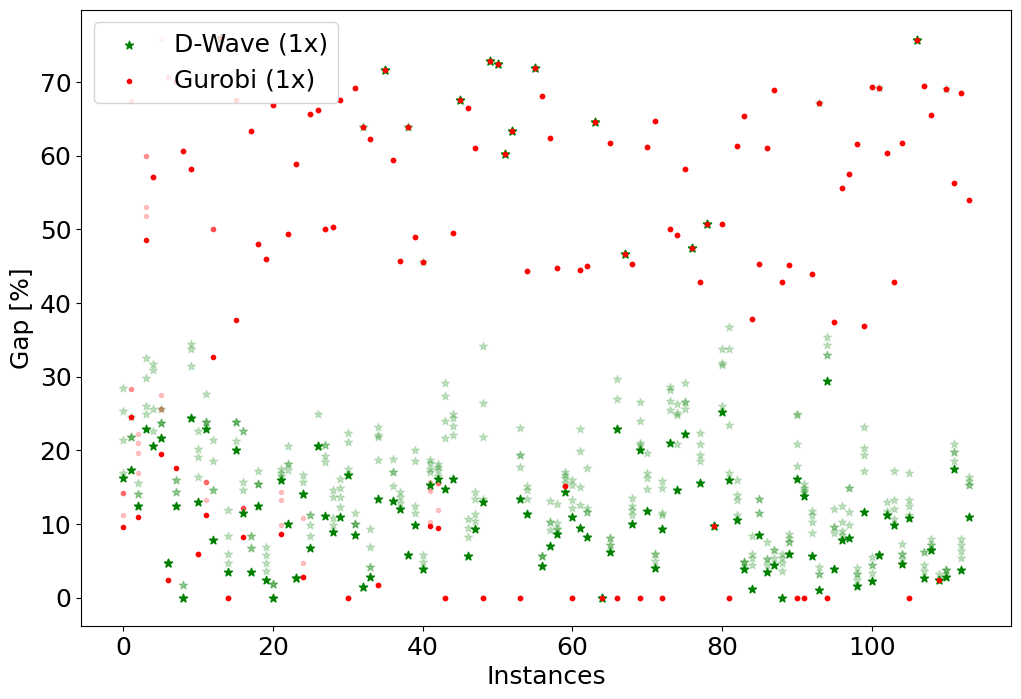}} 
    \subfigure{\includegraphics[width=0.49\textwidth]{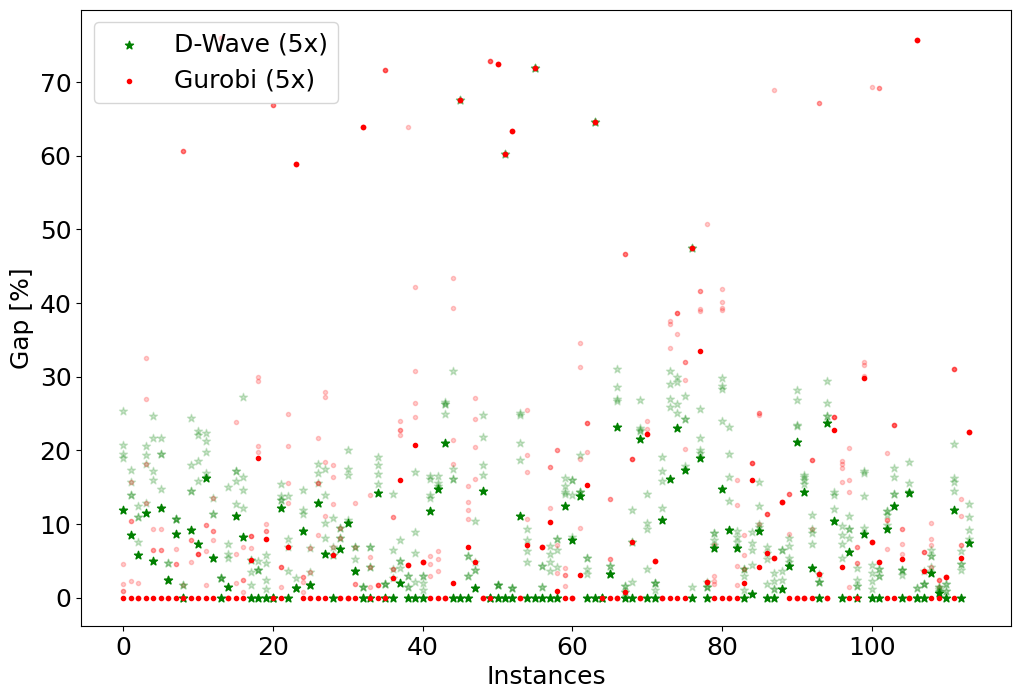}}
    \caption{The two figures compare the solution quality (gaps) of D-Wave and Gurobi for the QCBO formulation for every instance. The left for the 1x runtimes, the right for the 5x runtimes. The best runs are indicated by the dark stars and the dark dots, respectively.}
\label{fig:qcbo_gap}
\end{figure*}

In this section, we will compare the performance of Gurobi with D-Wave on the QCBO model. Figure \ref{fig:qcbo_gap} presents the gaps to the best solutions, while Table \ref{table_gap_qcbo_statistics} provides the statistical quantities. From Figure \ref{fig:qcbo_gap}, it is immediately apparent that D-Wave outperformed Gurobi for the 1x runtimes. Additionally, D-Wave maintained its superior performance across other runtimes, with further details available in Table \ref{table_gap_qcbo_wins}. The 5x runtimes present an intriguing case. At first glance, the figure might not definitively indicate D-Wave's superiority over Gurobi, as Gurobi demonstrates a very small gap for many of its best runs. Specifically, Gurobi achieves a gap smaller than $5\%$ for $68.42\%$ of instances, nearly matching the best solutions. In contrast, D-Wave only accomplishes this for $56.14\%$ of its runs. This discrepancy can be attributed to Gurobi's high variability, with numerous low gap runs and a few instances of very large gaps, as visible in the figure.

\begin{table}[ht]
\begin{center}
{\small
\begin{tabular}{ |c|c|c|c|c| } 
 \hline
 &  \multicolumn{2}{c|}{Gurobi} & \multicolumn{2}{c|}{D-Wave} \\
 \hline
 &  Average & \thead{Standard \\ deviation} & Average & \thead{Standard \\ deviation} \\ 
  \hline
 1x (best) & $42.46$ & $25.15$ & $16.61$ & $18.65$ \\ 
 2x (best) & $36.3$ & $26.11$ & $9.78$ & $15.17$\\  
 5x (best) & $9.67$ & $19.54$ & $5.68$ & $6.69$\\ 
  \hline
 1x (median) & $43.75$ & $24.84$ & $19.71$ & $18.62$ \\ 
 2x (median) & $38.59$ & $25.40$ & $15.07$ & $17.43$ \\ 
 5x (median) & $17.81$ & $21.50$ & $11.85$ & $13.24$ \\  
  \hline
 1x (worst) & $45.34$ & $24.64$ & $24.66$ & $19.08$ \\ 
 2x (worst) & $39.96$ & $24.88$ & $20.07$ & $19.21$ \\  
 5x (worst) & $23.85$ & $23.04$ & $14.71$ & $13.50$ \\ 
 \hline
\end{tabular}
}
\end{center}
\caption{The table compares the average gap and the standard deviation for Gurobi and D-Wave solving the QCBO model, for the best, median, and worst runs.}
\label{table_gap_qcbo_statistics}
\end{table}

\begin{table}[ht]
\begin{center}
{\small
\begin{tabular}{ |c|c|c|c| } 
 \hline
  &  1x & 2x & 5x \\ 
  \hline
 win Gurobi [$\%$] & $21.93$ & $22.81$ & $36.67$ \\ 
 win D-Wave [$\%$] & $62.63$ & $64.74$ & $55.96$ \\  
 \hline
\end{tabular}
}
\end{center}
\caption{The table shows how often D-Wave wins against Gurobi, and vica versa, on the QCBO for each of the three runtimes. }
\label{table_gap_qcbo_wins}
\end{table}

To fairly compare the time to solution, we must run Gurobi until it finds a solution with an objective value equal to or better than that of D-Wave. Our experiments showed that even after three hours, Gurobi often fails to match the solution quality achieved by D-Wave. Although some studies substitute maximum runtime for actual runtime, we have omitted such evaluations because this substitution would not provide meaningful insights.

\textit{Remark.} When solving the QCBO, Gurobi does not directly handle quadratic constraints. Instead, it linearizes the model into a MILP. A common linearization technique replaces each quadratic term $x_ix_j$ by the binary variable $y_{i,j}$ and the constraints $y_{i,j} \geq x_i + x_j - 1$ and $y_{i,j} \leq x_i, y_{i,j} \leq x_j$ to enforce $y_{i,j}=x_ix_j$. . However, this method has two main drawbacks. First, it introduces a quadratic number of new variables, which can lead to very large formulations. Second, the linear relaxation of this formulation is weak, often resulting in a larger branch-and-bound tree and consequently longer solution times.
\section{Conclusion}


\subsection{Performance}

From our experiments, we can rank the performance of the model solver pairs as follows:
\begin{itemize}
    \item[1)] MILP model with Gurobi,
    \item[2)] MILP model with D-Wave,
    \item[3)] QCBO model with D-Wave,
    \item[4)] QCBO model with Gurobi.
\end{itemize}
That is, the classical solver on the classical model is the most performant, and the classical solver on the alternative formulation is the least performant. 

The small amount of time spent on the QPU raises the question if quantum annealing has an effect at all on the optimization process and thereby the final solution.

\subsection{Modeling}
The general performance advantage of the MILP model over the QCBO model raises the question if the QUBO approach will lead to quantum advantage in the future, at least for D-Waves hybrid solvers. Currently, a very common way of doing quantum optimization is often to formulate a QUBO model and solve it with a quantum approach, for example, quantum annealing. Quantum annealing is the quantum component of the D-Wave quantum-classical metaheuristic and yet our QCBO, which is basically (almost) a QUBO\footnote{We did not move the quadratic constraints into the objective, which is a technicality, and did not convert the few remaining linear into a quadratic objective term.}, performed worse than the MILP model.

On the other hand, the performance difference for the 5x runtimes between solving the MILP with D-Wave and the QCBO is (quite) small, and for a couple of instances the QCBO model performed better. If these are the instances of interest, D-Wave benefits from a different modeling approach. In addition, solving QCBOs with quantum-classical metaheuristics is a young (research) field and significant advancement can be expected in three areas:
\begin{itemize}
    \item[1] Modeling the problem as a QCBO has not had any benefit until now. The standard way, which Gurobi uses, is to linearize such models. More research in modeling QCBOs can lead to more efficient and performant formulations.
    \item[2] The solution methods for QCBO models are not as advanced as for MILP models. Faster advantages can be expected.
    \item[3] The quantum hardware is still in its early stages. Advanced hardware can lead to better performance.
\end{itemize}

\subsection{Quadratic problems}

For solving our QCBO, we saw clear performance advantages when using D-Wave. Hence, problems that cannot be efficiently modeled as a MILP but as QCBO, can benefit from D-Wave already, similar to \cite{awasthi2024realworldapplicationquantumclassical}. Additionally, the QPLIB \cite{QPLIB} offers a promising starting point for further testing D-Wave's performance on QCBOs.


\section*{Acknowledgment}
We would like to acknowledge the Federal Ministry for Economic Affairs and Energy (abbreviated BMWE), for funding this work under the QCHALLenge Project (01MQ22008).

\bibliographystyle{ieeetr}
{\small
\bibliography{references}}

\begin{thebibliography}{10}

\bibitem{MoProLog}
{Frauenhofer IML}, ``{Modular Production Logistics},'' 2024.

\bibitem{awasthi2024realworldapplicationquantumclassical}
A.~Awasthi, N.~Kraus, F.~Krellner, and D.~Zambrano, ``{ Real World Application of Quantum-Classical Optimization for Production Scheduling },'' in {\em 2024 IEEE International Conference on Quantum Computing and Engineering (QCE)}, (Los Alamitos, CA, USA), pp.~239--244, IEEE Computer Society, Sept. 2024.

\bibitem{leibl}
D.~Leib, T.~Seidel, S.~J\"ager, R.~Heese, C.~Jones, A.~Awasthi, A.~Niederle, and M.~Bortz, ``An optimization case study for solving a transport robot scheduling problem on quantum-hybrid and quantum-inspired hardware,'' {\em Scientific Reports}, vol.~13, Oct. 2023.

\bibitem{comp_cqm_cplex}
A.~Bertuzzi, D.~Ferrari, A.~Manzalini, and M.~Amoretti, ``Evaluation of quantum and hybrid solvers for combinatorial optimization,'' in {\em Proceedings of the 21st ACM International Conference on Computing Frontiers}, CF '24, (New York, NY, USA), p.~232–239, Association for Computing Machinery, 2024.

\bibitem{portfolio_dwave_hybrid}
F.~Phillipson and H.~S. Bhatia, ``Portfolio optimisation using the d-wave quantum annealer,'' in {\em Computational Science -- ICCS 2021} (M.~Paszynski, D.~Kranzlm{\"u}ller, V.~V. Krzhizhanovskaya, J.~J. Dongarra, and P.~M.~A. Sloot, eds.), (Cham), pp.~45--59, Springer International Publishing, 2021.

\bibitem{Qgrid_demand_side}
D.~Bucher, J.~Nüßlein, C.~O'Meara, I.~Angelov, B.~Wimmer, K.~Ghosh, G.~Cortiana, and C.~Linnhoff-Popien, ``Incentivizing demand-side response through discount scheduling using hybrid quantum optimization,'' {\em IEEE Transactions on Quantum Engineering}, vol.~5, pp.~1--15, 2024.

\bibitem{SCHWORM202313}
P.~Schworm, X.~Wu, M.~Klar, J.~Gayer, M.~Glatt, and J.~C. Aurich, ``Resilience optimization in manufacturing systems using quantum annealing,'' {\em Manufacturing Letters}, vol.~36, pp.~13--17, 2023.

\bibitem{osaba2024hybridquantumsolversproduction}
E.~Osaba, E.~Villar-Rodr{\'i}guez, A.~Gomez-Tejedor, and I.~Oregi, ``Hybrid quantum solvers in production: How to succeed in the nisq era?,'' in {\em Intelligent Data Engineering and Automated Learning -- IDEAL 2024} (V.~Julian, D.~Camacho, H.~Yin, J.~M. Alberola, V.~B. Nogueira, P.~Novais, and A.~Tall{\'o}n-Ballesteros, eds.), (Cham), pp.~423--434, Springer Nature Switzerland, 2025.

\bibitem{Yarkoni_2022}
S.~Yarkoni, E.~Raponi, T.~Bäck, and S.~Schmitt, ``Quantum annealing for industry applications: introduction and review,'' {\em Reports on Progress in Physics}, vol.~85, p.~104001, sep 2022.

\bibitem{Schworm2023}
P.~Schworm, X.~Wu, M.~Glatt, and J.~C. Aurich, ``Solving flexible job shop scheduling problems in manufacturing with quantum annealing,'' {\em Production Engineering Research and Development}, vol.~17, p.~105–115, 2023.

\bibitem{Carugno2022}
C.~Carugno, M.~{Ferrari Dacrema}, and P.~Cremonesi, ``Evaluating the job shop scheduling problem on a d-wave quantum annealer,'' {\em Scientific Reports}, vol.~12, 2022.

\bibitem{SCHWORM2024142}
P.~Schworm, X.~Wu, M.~Klar, M.~Glatt, and J.~C. Aurich, ``Multi-objective quantum annealing approach for solving flexible job shop scheduling in manufacturing,'' {\em Journal of Manufacturing Systems}, vol.~72, pp.~142--153, 2024.

\bibitem{KU2016165}
W.-Y. Ku and J.~C. Beck, ``Mixed integer programming models for job shop scheduling: A computational analysis,'' {\em Computers \& Operations Research}, vol.~73, pp.~165--173, 2016.

\bibitem{gurobi}
{Gurobi Optimization, LLC}, ``{Gurobi Optimizer Reference Manual},'' 2023.

\bibitem{integer_programming}
M.~Conforti, G.~Cornuéjols, and G.~Zambelli, {\em Integer Programming}.
\newblock Springer Cham, 2016.

\bibitem{dwave_solver}
``D-wave hybrid solver service: An overview.''

\bibitem{farre}
P.~F. Catherine~McGeoch, ``Advantage processor overview,'' {\em D-Wave Technical Report Series}, 2022.

\bibitem{dwave_timer}
{D-Wave Systems Inc.}, ``dwave-system.'' \url{https://github.com/dwavesystems/dwave-system/blob/1.25.0/dwave/system/samplers/leap_hybrid_sampler.py#L768}, 2024.

\bibitem{QPLIB}
F.~Furini, E.~Traversi, P.~Belotti, A.~Frangioni, A.~Gleixner, N.~Gould, L.~Liberti, A.~Lodi, R.~Misener, H.~Mittelmann, N.~Sahinidis, S.~Vigerske, and A.~Wiegele, ``{QPLIB}: {A} library of quadratic programming instances,'' {\em Mathematical Programming Computation}, 2018.

\end{thebibliography}

\end{document}